\documentclass{svjour3}                     
\smartqed  
\usepackage{graphicx}
\usepackage{overpic}
\begin{document}

\title{Overview of $\bar K N$ and $\bar K$-nucleus dynamics}
\titlerunning{$\bar K$-nuclear dynamics} 

\author{Avraham Gal}

\authorrunning{A.~Gal} 

\institute{Avraham Gal \at
              Racah Institute of Physics, The Hebrew University,
Jerusalem~91904, Israel \\ \email{avragal@vms.huji.ac.il}}

\date{Received: date / Accepted: date}

\maketitle

\begin{abstract}
The main features of coupled-channel $\bar K N$ dynamics near threshold 
and its repercussions in few-body $\bar K$-nuclear systems are briefly 
reviewed highlighting the $I=1/2$ $\bar K NN$ system. For heavier nuclei, 
the extension of mean-field calculations to multi-$\bar K$ nuclear 
quasibound states is discussed focusing on kaon condensation. 

\end{abstract}

\section{Introduction}
\label{intro}

The $\bar K$-nucleus interaction near threshold is strongly attractive 
and absorptive as suggested by fits to the strong-interaction shifts 
and widths of $K^-$-atom levels~\cite{BFG97,FG07}. Global fits yield 
extremely deep density dependent optical potentials with nuclear-matter 
depth Re$V_{\bar K}(\rho_0)\sim-$(150-200) MeV at threshold. Chirally 
based coupled-channel models that fit the low-energy $K^-p$ reaction 
data, and the $\pi\Sigma$ spectral shape of the $\Lambda(1405)$ 
resonance, yield moderate depths Re$V_{\bar K}(\rho_0)\sim-$100 MeV, 
as summarized recently in Ref.~\cite{WH08}. A major uncertainty in these 
chirally based studies arises from fitting the $\Lambda(1405)$ resonance 
by the imaginary part of the $\pi\Sigma(I=0)$ amplitude calculated within 
the same coupled channels chiral scheme. A third class, of shallower 
potentials with Re$V_{\bar K}(\rho_0) \sim -$(40-60) MeV, was obtained 
by imposing a Watson-like self-consistency requirement \cite{RO00}. 
However, one needs then to worry about higher orders in the chiral 
expansion which are not yet in. 

I start by making introductory remarks on the $\bar KN - \pi\Sigma$ system, 
followed by reviewing two topics related to $\bar K$ nuclear quasibound 
states: (i) the $K^-pp$ system as a prototype of few-nucleon quasibound 
states of $\bar K$ mesons; and (ii) multi-$\bar K$ nucleus quasibound states. 
In reviewing the latter topic I will discuss the phenomenological evidence 
for the `extremely deep' $\bar K$-nucleus potentials used in nuclear 
and nuclear-matter calculations. 

\section{Polology of $\bar KN - \pi\Sigma$ coupled channels} 
\label{sec:poles} 

\begin{figure}[t]
\centering
\begin{tabular}{cp{1cm}c}
\begin{overpic}[height=6cm,width=0.54\textwidth,clip]{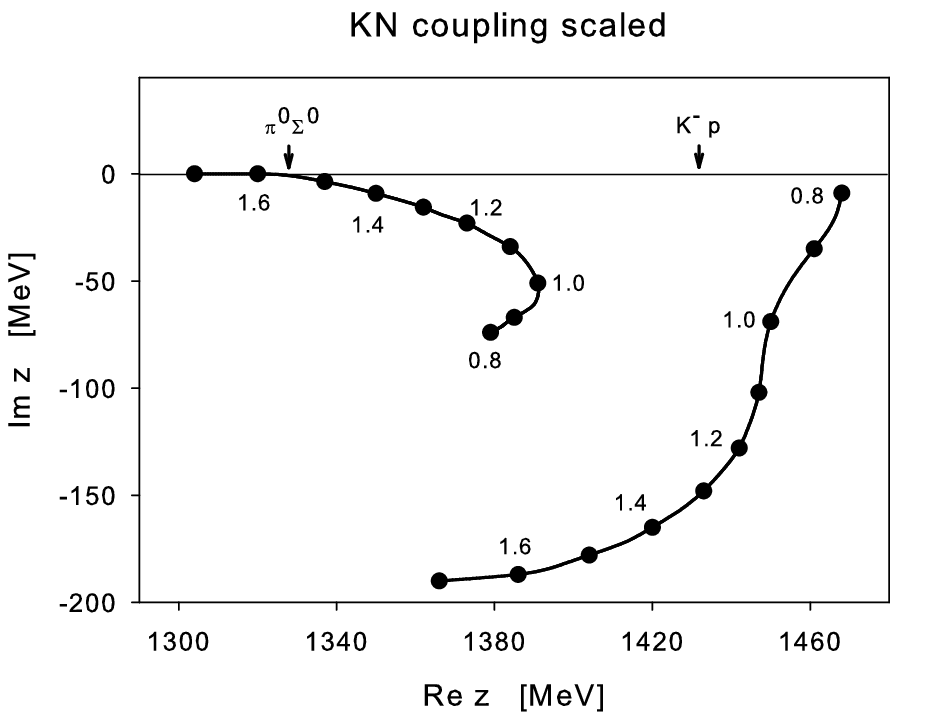}
\end{overpic}
\begin{overpic}[height=6cm,width=0.44\textwidth,clip]{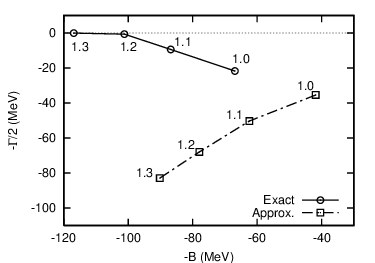}
\end{overpic}
\end{tabular}
\caption{Left: trajectories of Gamow poles in the complex energy (z) plane, 
on the Riemann sheet [$\Im k_{\bar K N},\Im k_{\pi \Sigma}$] = [$+,-$], upon 
scaling the $\bar K N$ interaction strengths (taken from Ref.~\cite{CG08}). 
The $\pi^0\Sigma^0$ and $K^-p$ thresholds are marked by arrows. 
Right: $\bar K NN(I=1/2)$ quasibound state energy from Ref.~\cite{IS08} as 
a function of the $\bar K N$ interaction strength within a three-body coupled 
channel calculation (circles) and within a single channel approximate 
calculation (squares).} 
\label{fig:1}
\end{figure}

Modern chirally motivated $\bar KN - \pi\Sigma$ coupled-channel models 
give rise to {\it two} Gamow poles that dominate low-energy $\bar K N$ 
dynamics. Representative pole positions are shown on the left-hand side 
of Fig.~\ref{fig:1} for the coupled channels model of Ref.~\cite{CS07}, 
together with the trajectories followed by these poles upon scaling the 
$\bar K N$ interactions. This model fits well all the low-energy $K^-p$ 
scattering and reaction data. It reproduces reasonably well the $\pi\Sigma$ 
spectrum shape, identified with the $\Lambda(1405)$ $\pi\Sigma$ resonance, 
which is determined primarily by the lower pole at $(1391,-{\rm i}51)$ MeV. 
This identification is further supported by the trajectory of the lower pole 
which merges into an $I=0$ genuinely bound state below the $\pi^0\Sigma^0$ 
threshold when the $\bar K N$ interactions are sufficiently increased. 
The upper pole, in this model, is located above the $K^-p$ threshold. 
However, its position and the trajectory it follows away from the real 
energy axis are model dependent and sensitive to off-shell 
effects.\footnote{For example, the pole positions in Ref.~\cite{HW08} are 
$z_>=1428-{\rm i}17,~z_<=1400-{\rm i}76$ MeV.} 
As discussed below in Sect.~\ref{sec:KNN}, the upper pole affects 
significantly the three-body $[\bar K (NN)_{I=1} - \pi\Sigma N]_{I=1/2}$ 
dynamics of the $K^-pp$ system. The energy and width of the 
($\bar K NN$ quasibound~-~$\pi\Sigma N$ resonannce) state are determined by 
a Gamow pole whose trajectory, from Ref.~\cite{IS08}, is depicted in circles 
on the right-hand side of Fig.~\ref{fig:1}. Similarly to the lower-pole 
$\Lambda(1405)$ trajectory in the two-body case, this three-body pole also 
merges below the $\pi\Sigma N$ threshold into a genuinely bound state which, 
upon extending the model space, becomes a quasibound $\pi\Sigma N$ state 
decaying to lower channels ignored here.\footnote{The other trajectory, 
depicted in squares, is relevant only to the discussion in 
Sect.~\ref{sec:KNN}.}  

\section{Few-nucleon $\bar K$ systems} 
\label{sec:KNN}

The lightest $\bar K$ nuclear configuration maximizing the strongly attractive 
$I=0~\bar K N$ interaction is $[\bar K (NN)_{I=1}]_{I=1/2,J^{\pi}=0^-}$, 
loosely denoted as $K^-pp$. The FINUDA collaboration presented evidence in 
$K^-$ stopped reactions on several nuclear targets for the process $K^-pp \to 
\Lambda p$, interpreting the observed signal as due to a $K^-pp$ deeply bound 
state with $(B, \Gamma) \approx (115, 67)$~MeV \cite{ABB05}. However, this 
interpretation has been challenged in Refs.~\cite{MFG06,MOR06}. A preliminary 
new analysis of DISTO $pp \to K^+ \Lambda p$ data was presented in EXA08 
suggesting a $K^-pp$ signal with $(B, \Gamma)\approx (105, 118)$~MeV 
\cite{Yam08}. The location practically on top of the $\pi\Sigma N$ threshold, 
and particularly the large width, are at odds with any of the few-body 
calculations listed below, posing a problem for a $K^-pp$ quasibound state 
interpretation. 

\begin{figure} 
\includegraphics[height=6.0cm,width=1.0\textwidth]{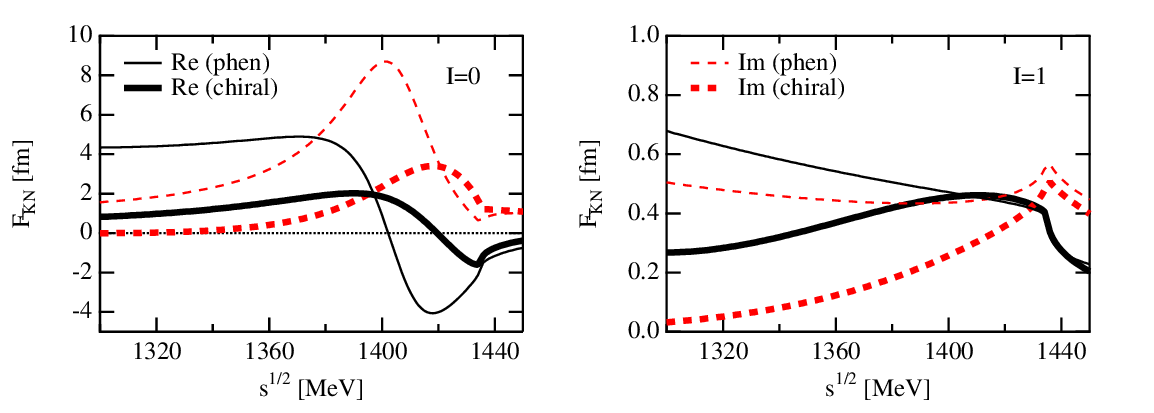} 
\caption{Comparison of the Yamazaki-Akaishi phenomenological $\bar K N$ 
amplitudes \cite{AY02} with the Hyodo-Weise chirally based $\bar K N$ 
amplitudes \cite{HW08}. Figure taken from Ref.~\cite{HW08}.} 
\label{fig:2}
\end{figure} 

\begin{table} 
\caption{Calculated $B_{K^-pp}$, mesonic ($\Gamma_{\rm m}$) 
\& nonmesonic ($\Gamma_{\rm nm}$) widths (in MeV) of $K^-pp$.} 
\label{tab:1} 
\begin{tabular}{llllll} 
\hline\noalign{\smallskip}
&\multicolumn{2}{c}{$\bar K NN$ single channel} 
&\multicolumn{3}{c}{$\bar K NN - \pi\Sigma N$ coupled channels} \\ 
& ATMS~\cite{YA02,AY02} & AMD~\cite{DHW08} & Faddeev~\cite{SGM07} & 
Faddeev~\cite{IS07} & variational~\cite{WG08} \\ 
\noalign{\smallskip}\hline\noalign{\smallskip}
$B_{K^-pp}$ & ~~~48 & ~17--23 & ~~50--70 & ~~60--95 & ~~~40--80 \\ 
$\Gamma_{\rm m}$ & ~~~61 & ~40--70  & ~~90--110 & ~~45--80 & ~~~40--85 \\ 
$\Gamma_{\rm nm}$ & ~~~12 & ~~4--12 & & & ~~~$\sim 20$ \\
\noalign{\smallskip}\hline
\end{tabular}
\end{table}

Results of few-body calculations for the $K^-pp$ system are displayed in 
Table~\ref{tab:1}. The marked difference between the `$\bar K NN$ single 
channel' binding energies $B_{K^-pp}$ reflects the difference between the 
input $\bar K N$ amplitudes shown in Fig.~\ref{fig:2}: the Yamazaki-Akaishi 
$I=0$ single-pole amplitude \cite{YA02} resonates at 1405 MeV, 
whereas the Dote-Hyodo-Weise $I=0$ amplitude \cite{DHW08} resonates at 
1420 MeV (close to the upper of two poles). This dependence on the input 
amplitudes has been verified in coupled-channel Faddeev calculations 
\cite{IS08,SGMR07} and in variational calculations \cite{WG08}. 

A notable feature of the $K^-pp$ coupled-channel calculations 
\cite{SGM07,IS07,WG08} in Table~\ref{tab:1} is that the explicit use of 
the $\pi\Sigma N$ channel adds about $20 \pm 5$~MeV to the binding energy 
calculated using effective $\bar K N$ potential within a single-channel 
calculation. This is demonstrated on the right-hand side of Fig.~\ref{fig:1} 
by comparing corresponding points on the two trajectories shown there.

\section{$\bar K$-nucleus potentials from kaonic atoms and from nuclear 
reactions} 
\label{sec:katoms}  

\begin{figure}[t]
\centering
\begin{tabular}{cp{1cm}c}
\begin{overpic}[height=6cm,width=0.44\textwidth,clip]{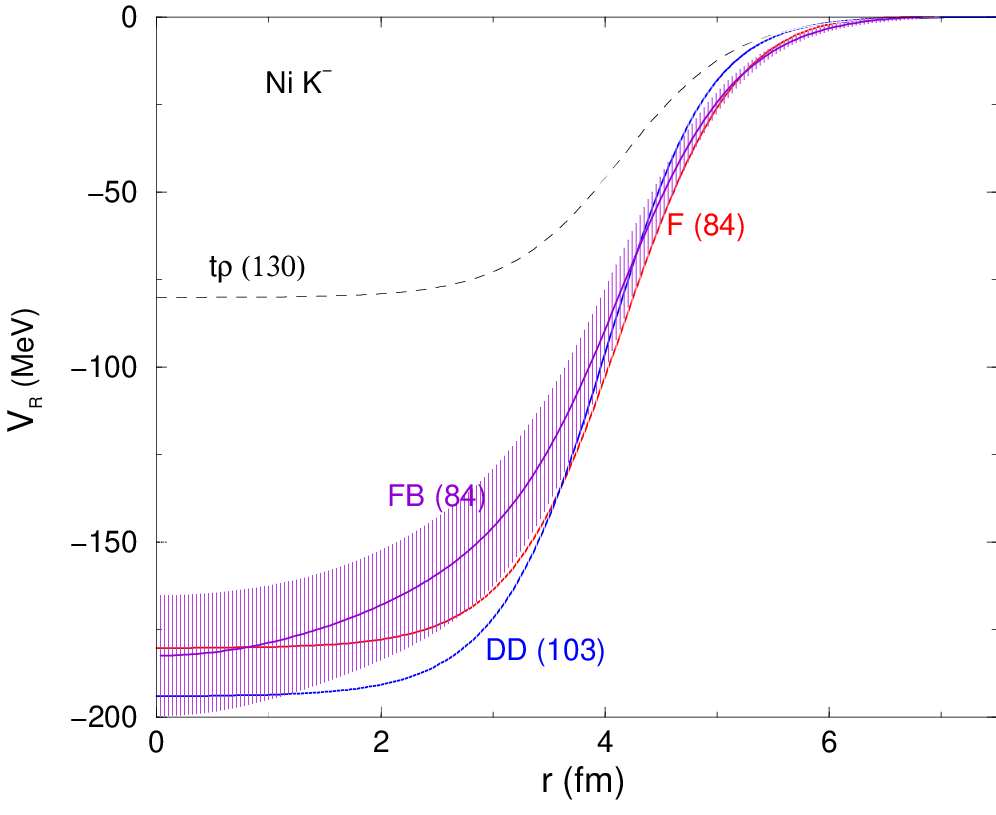}
\end{overpic}
\begin{overpic}[height=6cm,width=0.48\textwidth,clip]{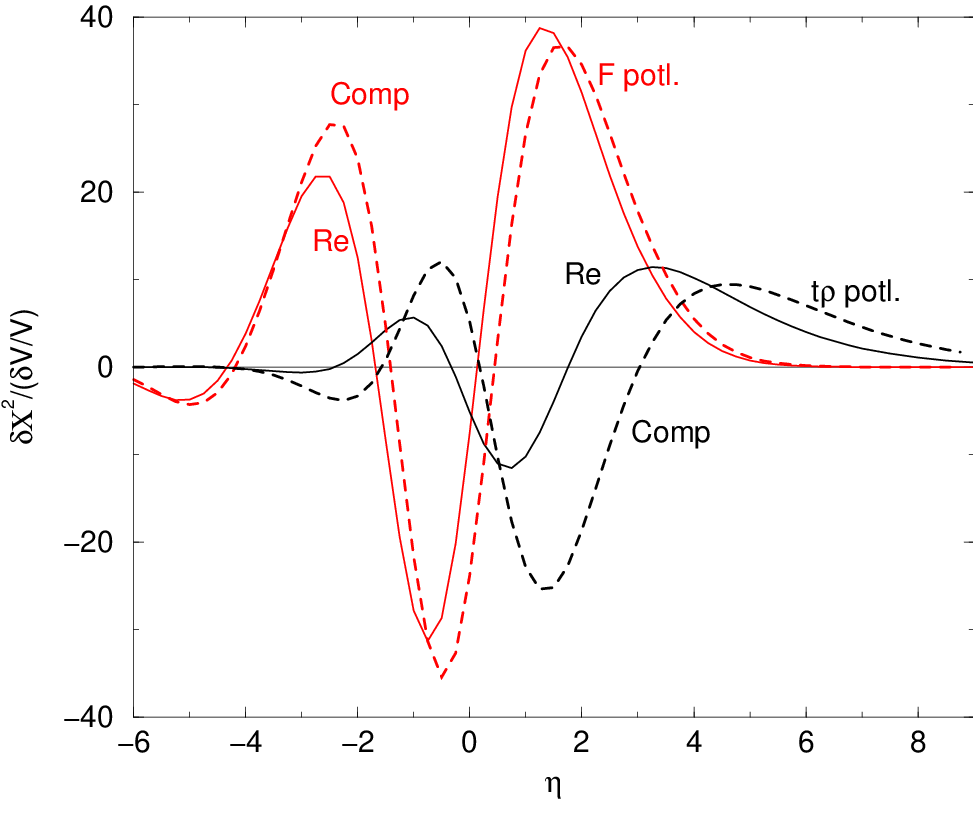}
\end{overpic}
\end{tabular}
\caption{Comparisons between density dependent potentials (DD, FB, F) 
and a $t\rho$ potential fitted to kaonic-atom data~\cite{FG07}. 
Left: the real part of the $\bar K - {^{58}{\rm Ni}}$ potential. 
Right: functional derivatives of kaonic atoms $\chi^2$ with respect 
to the fully complex (Comp, dashed) and real (Re, solid) potential 
as a function of $\eta=(r-R_c)/a_c$ using 2pF charge density distributions.}
\label{fig:3}
\end{figure}

Figure \ref{fig:3} (left) illustrates the real part of the best-fit 
$\bar K$-nucleus potential for $^{58}$Ni as obtained for several models. 
The corresponding values of $\chi ^2$ for 65 $K^-$-atom data points are 
given in parentheses. A Fourier-Bessel (FB) fit \cite{BF07} is also shown, 
within an error band. Just three terms in the FB series, added to a $t\rho $ 
potential, suffice to achieve a $\chi ^2$ as low as 84 and to make the 
potential extremely deep, in agreement with the density-dependent best-fit 
potentials DD and F. In particular, the density dependence of potential F 
provides by far the best fit ever reported for any global $K^-$-atom data fit, 
and the lowest $\chi ^2$ value as reached by the model-independent FB method. 

The functional derivative (FD) method for identifying the radial regions to 
which exotic atom data are sensitive is demonstrated in Fig.~\ref{fig:3} 
(right) for the F and $t \rho $ best-fit potentials \cite{BF07}. It is clear 
that whereas within the $t\rho$ potential there is no sensitivity to the 
interior of the nucleus, the opposite holds for the density dependent F 
potential which accesses regions of full nuclear density. This owes partly 
to the smaller imaginary part of F, which also explains why the FD for the 
complex F potential is well approximated by that for its real part. 

\begin{figure}[t]
\centering
\begin{tabular}{cp{1cm}c}
\begin{overpic}[height=6cm,width=0.48\textwidth,clip]{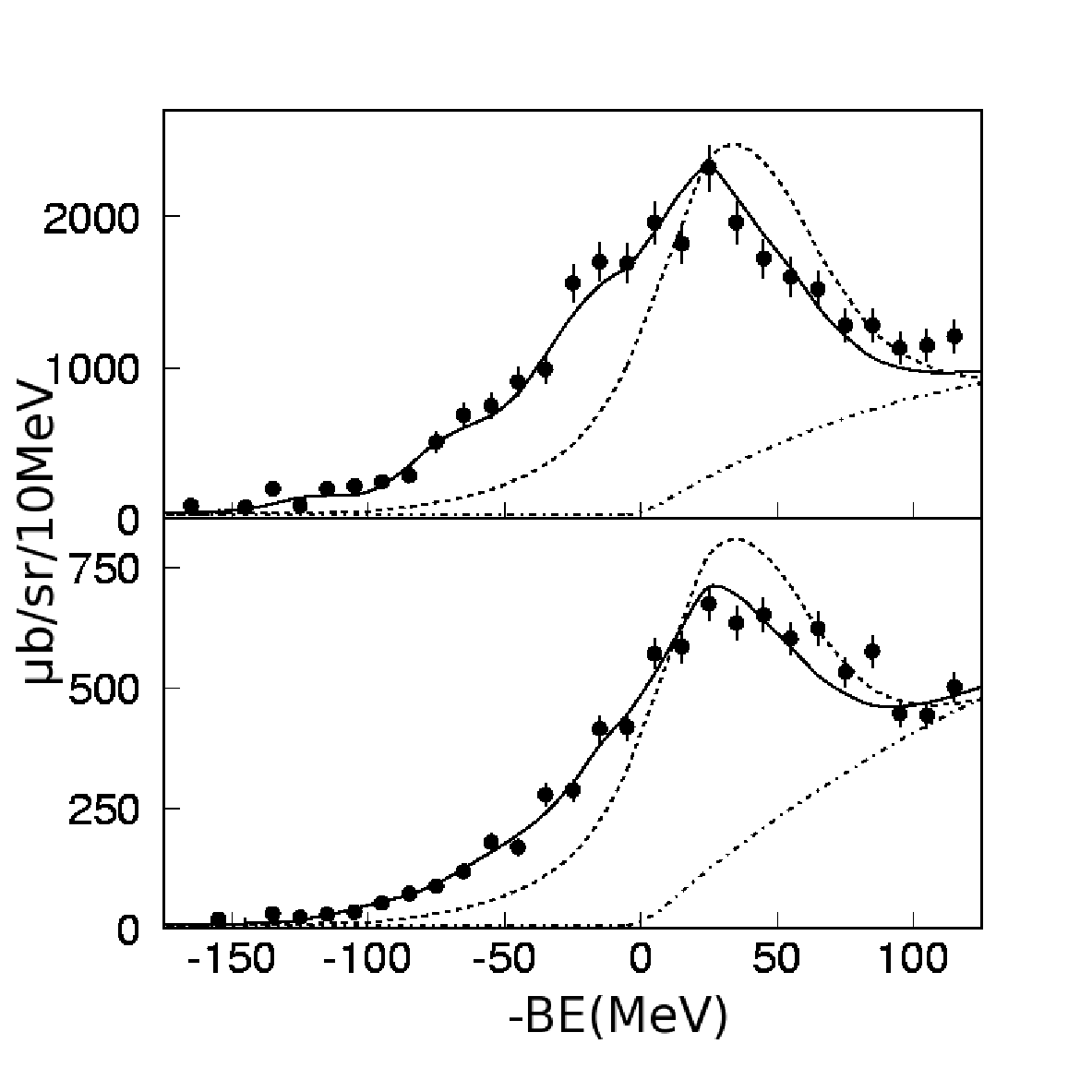}
\end{overpic}
\begin{overpic}[height=6cm,width=0.50\textwidth,clip]{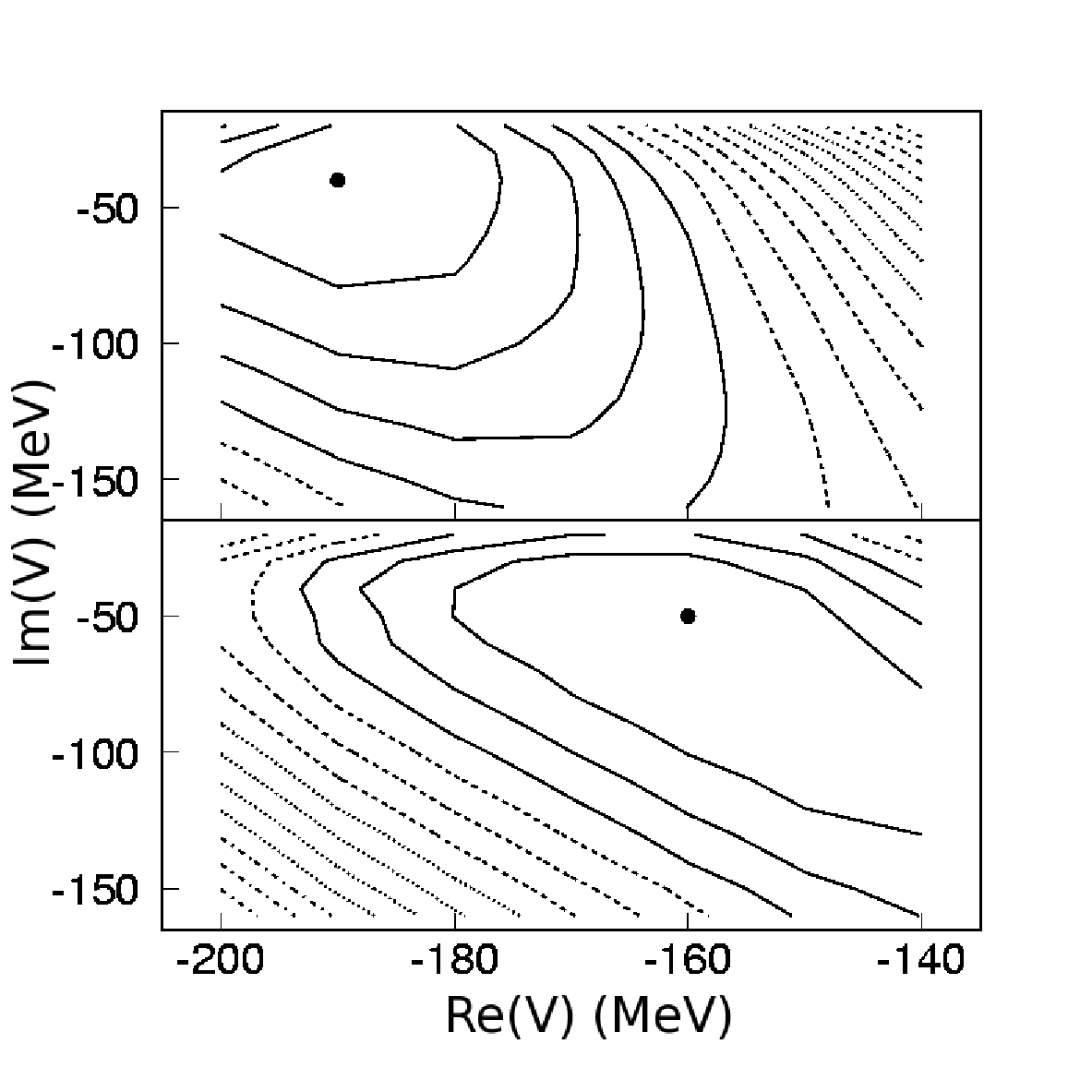}
\end{overpic}
\end{tabular}
\caption{KEK-PS E548 missing mass spectra (left) and $\chi^2$ contour 
plots (right) for $(K^-,n)$ (upper) \& $(K^-,p)$ (lower) at 
$p_{K^-}=1$~GeV/c on $^{12}$C \cite{kish07}.}
\label{fig:4}
\end{figure}

A fairly new and independent evidence in favor of extremely deep 
$\bar K$-nucleus potentials is provided by $(K^-,n)$ and $(K^-,p)$ 
spectra taken at KEK on $^{12}$C \cite{kish07} and very recently also 
on $^{16}$O (presented in PANIC08) at $p_{K^-}=1$ GeV/c. The $^{12}$C 
spectra are shown in Fig.~\ref{fig:4}, where the solid lines on the 
left-hand side represent calculations (outlined in Ref.~\cite{YNH06}) 
using potential depths in the range 160-190 MeV. The dashed lines 
correspond to using relatively shallow potentials of depth about 60 MeV 
which I consider therefore excluded by these data.  

In conclusion, optical potentials derived from the observed 
strong-interaction effects in kaonic atoms and from $(K^-,N)$ nuclear 
spectra are sufficiently deep to support strongly-bound antikaon states. 
However, a fairly sizable extrapolation is required to argue for 
$\bar K$-nuclear quasibound states at energies of order 100 MeV below 
threshold, using a potential determined largely near threshold. 

\section{Multi-$\bar K$ nucleus quasibound states from RMF calculations} 
\label{sec:RMF}

\begin{figure}[t]
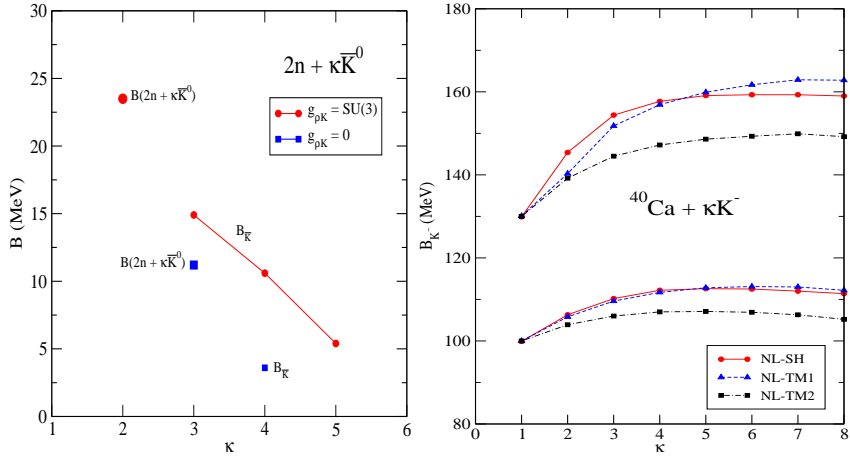

\centering
\begin{tabular}{cp{1cm}c}
\begin{overpic}[height=6cm,width=0.44\textwidth,clip]{2nk.eps}
\end{overpic}
\begin{overpic}[height=6cm,width=0.47\textwidth,clip]{multik40ca.eps}
\end{overpic}
\end{tabular}
\caption{RMF calculations of multi-$\bar K$ nucleus quasibound states as 
function of the number $\kappa$ of $\bar K$ mesons. Left: for two neutrons, 
demonstrating the isovector effect. Right: for $^{40}{\rm Ca}$ core, for 
several nuclear RMF models, with two choices of parameters fixed 
for $\kappa=1$ \cite{GFGM08}.}
\label{fig:5}
\end{figure}

Relativistic mean field (RMF) calculations of single- and of multi-$\bar K$ 
nuclei are reported in these Proceedings by J. Mare\v{s}. Dynamical 
calculations of single-$\bar K$ medium and heavy nuclei produce 
quasibound states bound by 100-150 MeV for potentials compatible with 
$K^-$ atom data. These calculations also provide a quantitative estimate 
of the expected widths, which are larger than 100 MeV near threshold and 
remain of order 50 MeV or more, even as the primary $\bar K N \to \pi\Sigma$ 
decay mode shuts off at about 100 MeV below threshold \cite{MFG06,GFGM07}. 
Highlights of multi-$\bar K$ nuclear calculations are demonstrated here in 
Fig~\ref{fig:5}. On the left-hand side, results of RMF calculations are shown 
for $2n+\kappa{\bar K}^0$ systems, where all decay channels are suppressed. 
For $\kappa=1$, the ${\bar K}^0 nn$ system which is charge symmetric to 
$K^-pp$ was found to be unbound, apparently because RMF calculations do not 
allow for a $\bar K N - \pi\Sigma$ channel coupling. Binding within these 
schematic calculations starts at $\kappa=2$ if isovector degrees of freedom 
are treated properly (say, using SU(3)) and for $\kappa=3$ if they are 
suppressed. The ${\bar K}^0$ separation energy, denoted $B_{\bar K}$, is 
found to decrease with $\kappa$ which is a special case of the saturation 
property established in heavier system, as discussed below. 

$K^-$ separation energies $B_{K^-}$ in multi-$K^-$ nuclei 
$^{40}{\rm Ca}+\kappa K^-$ are shown on the right-hand side of 
Fig.~\ref{fig:5} for two choices of $g_{\sigma K}$, designed within each 
RMF model to produce $B_{K^-}=100$ and $130$ MeV for $\kappa=1$. 
A robust saturation of $B_{K^-}$ with $\kappa$, independently of the 
applied RMF model, emerges from these calculations. 
The saturation values of $B_{K^-}$ do not allow conversion of $\Lambda$ 
hyperons to $\bar K$ mesons through strong decays $\Lambda \to p + K^-$ 
or $\Xi^- \to \Lambda + K^-$ in multi-strange hypernuclei, which therefore 
remain the lowest-energy configuration for multi-strange systems. This 
provides a powerful argument against $\bar K$ condensation in the laboratory, 
under strong-interaction equilibrium conditions \cite{GFGM08,GFGM09}. 
It does not apply to kaon condensation in neutron stars, where equilibrium 
configurations are determined by weak-interaction conditions.  

\begin{acknowledgements}
I am indebted to my collaborators Ale\v{s} Ciepl\'{y}, Eli Friedman, 
Daniel Gazda and Ji\v{r}\'{i} Mare\v{s} for instructive discussions 
and fruitful cooperation. This research was supported in part by the 
Israel Science Foundation, Jerusalem, grant 757/05.
\end{acknowledgements}

\end{document}